\documentclass[11pt, a4paper]{article}
\pdfoutput = 1
\usepackage{graphicx}
\usepackage{amsmath}
\begin{document}
 \numberwithin{equation}{section}

\title{Thermal duality and gravitational collapse in heterotic string theories.}
\author{Michael Hewitt \\
Department of Computing, Canterbury Christ Church University,\\
North Holmes Road, Canterbury, CT1 1QU, U.K. \\ 
 }
\date{4 July 2015}
\maketitle
\begin{abstract}
 \
The thermal duality of  $E(8)$ x $E(8)$ and $SO(32)$  heterotic string theories may underpin a mechanism that would convert the kinetic energy of infalling matter during gravitational collapse to form a region of a hot string phase that would expel gravitational gradients. This phase would be the continuation of a Ginzburg-Landau like superconductor in the Euclidean regime. In this scenario, there would be no event horizon or singularity produced in gravitational collapse. Solutions are presented for excitations of the string vacuum that may form during gravitational collapse and drive the transition to the hot phase. The proposed mechanism is developed here for the case of approximately spherical gravitational collapse in 4 uncompactified spacetime dimensions. A way to reconcile the large entropy apparently produced in this process with quantum mechanics is briefly discussed. In this scenario, astrophysical objects such as stellar or galactic cores which have undergone extreme gravitational collapse would have converted to regions of this hot string phase. The relationship of this proposal to the `firewall paradox' is noted. \\
\


\end{abstract}
\
\

\section{Introduction}

\    \ Black holes \cite{Wald} are generally accepted as possible endpoints of gravitational collapse. Oppenheimer and Snyder predicted the formation of black holes in some stellar collapses \cite{OppenheimerSnyder}, and for some astrophysical situations there is no generally accepted mechanism to prevent the formation of a black hole.  Indeed there are strong candidate objects to have reached this stage of compression among supernova remnants \cite{Orosz} and galactic cores \cite{Bender}.  We wish to propose that there is a mechanism in the $E(8)$ x $E(8)$ and $SO(32)$ heterotic string theories which would avert horizon formation in all cases where no other mechanism comes into play, with regions of a hot string phase forming instead of black holes. The scenario which is proposed would also resolve the recently noted `firewall paradox' \cite{Almheiri}, \cite{Braunstein} \footnote{The author thanks Sam Braunstein for bringing \cite{Braunstein} to his attention.} in a radical way. Here, quantum mechanics and the equivalence principle are both respected, but the black hole complementarity principle \cite{Susskind} is abandoned. An infalling observer would indeed encounter a kind of `firewall' as novel processes rooted in thermal duality are triggered. \\

The process of black hole formation in general relativity was shown to be generic in the Penrose singularity theorem \cite{Penrose}.  The scenario in general relativity is summarised by a Penrose diagram with all world lines which cross the horizon meeting a spacelike singularity set within a finite time. Penrose identified \cite{Penrose} a point of no return after which a singular endpoint cannot be averted (except possibly by invoking negative energy) as the formation of a closed trapped surface \cite{HawkingEllis}. It is just prior to the formation of such trapped surfaces that the mechanism proposed in this paper would come into play, circumventing the formation of horizons. The causal structure of spacetime would then be isomorphic to that of Minkowski space.\\

The necessity of a singularity within a black hole can be understood as follows.  If we look at a forming black hole as a particle accelerator, suppose that two particles (e.g. electrons) falling in from opposite sides could collide. At the collision, since both particles have timelike worldlines, the centre of mass energy of the two particles would take some finite value $E$.  However, in the reference frame of an observer at the centre of mass, the kinetic energy of the particles will already have exceeded $E$ (and indeed any finite value) during the collapse.  To avoid the possibility of such a contradiction, the causal structure of spacetime must make any such encounter impossible, and this is realised by the presence of the singularity which enforces a spacelike separation of the particle worldlines to their ends.\\

Towards the singularity, an unboundedly large energy is compressed within an unboundedly small volume \cite{Penrose}.  A natural conjecture would be that high energy effects (perhaps related to string theory) would intervene in this state of high compression to resolve the singularity, as the conditions are reminiscent of a time-reversed version of those prevailing in the early universe.  However, the mechanism which is thought to resolve the singularity at the beginning of time predicted by the Hawking singularity theorem \cite{HawkingEllis}, namely inflation, relies on negative pressure filling boundaryless spatial hypersurfaces.  As the collapsing object does not fill space, the negative pressure needed to support rapid deflation would have no support at the object's surface.  The best strategy for averting a singularity in the case of collapse thus seems to be to prevent the formation of a closed horizon.\\
 
Can quantum mechanics, and in particular string theory provide such a mechanism?  Near the horizon the temperature apparently diverges for a static observer held in place at a fixed distance from the horizon, due to the Unruh effect \cite{Unruh}.  However, string theory apparently implies that there is a maximum temperature (the Hagedorn temperature \cite{Hagedorn}) suggesting that horizon formation may be averted by string effects \cite{Hewitt93} or be taken to be equivalent to a string theoretic picture \cite{Susskind}.  In themselves, however, these observations do not provide the mechanism for braking the fall of matter during gravitational collapse which would be needed to avert horizon formation. \\

String and brane state counting for near critical holes \cite{StromingerVafa}, \cite{Maldacena} in type II string theory gives agreement with the Bekenstein-Hawking entropy \cite{Hawking76}, and can be pictured as being due to string or D brane degrees of freedom near the horizon \cite{Maldacena}.  The analyses of \cite{StromingerVafa}, \cite{Maldacena} rely on counting D brane degrees of freedom, which are absent in the $E(8)$ x $E(8)$  heterotic string regime \cite{Grossetal}, due to the absence here of D branes and the unbreakable nature of E8 strings for topological reasons \cite{Polchinski}.  The suggested string-hole equivalence would thus have to be realised in a different way for $ E(8)$ x $E(8)$ strings. The possibility of such a connection seems worth studying for two reasons:  the string-hole correspondence seems likely for reasons of consistency to hold for all regions of M theory, including $E(8)$ x $E(8)$ strings, and more pragmatically because this kind of string model is more likely to be relevant to astrophysical gravitational collapse. \\

A closer analysis of the Hagedorn transition \cite{AtickWitten}, \cite{Sathiapalan}, \cite{ObrienTan} indicates the possibility of a high temperature string phase for heterotic strings characterised by the formation of a condensate. The heterotic nature of the strings is essential to this mechanism.   This phase was first noted for its possible relevance to gravitational collapse in \cite{Hewitt93}. In \cite{Hewitt02} it was suggested that the endpoint of gravitational collapse would be a region of the high temperature phase, having a hyperbolic internal spatial geometry, allowing the entropy to be reconciled with the Hawking value, which is related to the surface area rather than the Euclidean volume.  A non-local quantum mechanical mechanism which would allow conversion of infalling matter to the high temperature string phase through the spontaneous formation of shells of the high temperature string phase was also proposed in \cite{Hewitt02}.  It was conjectured that the nucleation of such shells would take place where the surface gravity reaches the duality temperature, giving a close relationship between nucleation sites and proximity to the point where Penrose surfaces would otherwise form.\\

 The purpose of this communication is to replace the shell nucleation mechanism of \cite{Hewitt02}, which relies on no-local quantum correlations around the collapsing object with a process that works locally to convert the kinetic energy of infalling matter into thermal excitations of the string vacuum.   The conditions for thermalon excitation during an astrophysical gravitational collapse are identified, and the mechanism by which the kinetic energy of infalling matter is converted to the thermalized energy of the high temperature string phase is elucidated.  It will be shown that the conditions under which the conversion of energy to thermalon excitations and thence to high temperature string phase would occur only in final stages of gravitational collapse, and that the normal vacuum would be stable against such processes.  A dissipative mechanism for the thermalisation of the kinetic energy of infalling matter is proposed, which could lead to a final state almost in thermal equilibrium with the surrounding vacuum (with only Hawking like radiation escaping) and which would not have an event horizon. A possible relationship between thermodynamic thermalon entropy and an entropy of entanglement between short and long string sectors will be proposed which could enable the generation of the Hawking entropy to be reconciled with conventional quantum mechanical state evolution.\\

The thermal equilibrium between the high temperature string phase and normal vacuum indicates that the entropy of the high temperature string phase should agree with the Hawking value to leading order in $ M/M_p$ ($M_p$ is the Planck mass).  It thus seems that a microscopic model of the Hawking entropy may also be possible for $E(8)$ x $E(8)$ strings.  The approach taken in this paper addresses the issue of string-hole correspondence in a different regime and with different methods to \cite{StromingerVafa}, \cite{Maldacena}. The latter invoke supersymmetry to carry out an evaluation of the number of microstates by direct enumeration, in a way which is independent of thermodynamic arguments. This takes place in a critical limit, where the hole becomes cold. In contrast, the current paper relies on properties of the heterotic thermalon, which intrinsically breaks supersymmetry – supersymmetry is necessarily broken in a thermal ensemble, and indeed the thermalon does not even belong to a supersymmetry multiplet.  Agreement with the Hawking entropy value follows directly to leading order from the thermodynamic equilibrium between the hot phase and the ambient normal vacuum, although the counting of string microstates can only be inferred  at this stage.  Again in contrast to the cases considered in \cite{StromingerVafa}, \cite{Maldacena} it is the approximately Schwarzschild case which is easiest to analyse for the thermalon mechanism, rather than the critical ones. The justification of the Hawking entropy in such contrasting scenarios seems to support the conjecture that it should be valid for all M theory cases. \\

\

\section{Thermalons}

The partition sum $Z$ at a finite temperature may be calculated by compactifying time in an imaginary direction. This introduces a winding mode in string theory \cite{AtickWitten}, which we will refer to as the thermalon. This situation may be represented by a compactification of the string target space on a Lorentzian lattice. A change of $\beta$ is represented by a Lorentzian transformation \cite{AtickWitten}. The critical points at which the thermalon becomes a tachyon are located where the Lorentzian trajectory crosses the circle in the $(L,R)$ energy plane representing zero mass, and this gives a value for the Hagedorn temperature \cite{Hagedorn} which agrees with that found by counting string states. For free strings, the partition sum $Z$ for a single string diverges at this temperature. Although the thermalon is, in itself, a fictitious particle it provides an effective theory which summarises the overall effect of the tower of excited string states. When the continuation is made back to the physical Minkowski regime, the interaction of the thermalon field with the background geometry presented below is intended to provide an effective theory which summarises interactions of strings with a condensate of string tower states. For interacting strings, the Hagedorn temperature is marked by the formation of this condensate.\\

For heterotic strings, of either the $E(8)$ x $E(8)$ or $SO(32)$ types, the winding and momentum numbers $(w,p)$  of the thermalon state which may become a tachyon are $(1,1)$. This is possible because the GSO projection is reversed for consistency for the winding state  \cite{AtickWitten}.  Note that the combination of wrapping and momentum quantum numbers for the thermalon around the imaginary time circle is the same as that for left moving heterotic modes. There is a contrast between the Lorentzian trajectories for the thermalon in the non-heterotic and heterotic cases, with the heterotic  case exhibiting thermal duality. The thermalon is a tachyon where its trajectory lies inside the circle (in the Euclidean norm of conformal weights) representing zero mass.  In the non-heterotic case, the trajectory follows a straight line towards the origin, and the thermalon becomes more tachyonic as the temperature is increased further (see Figure 1).\\
\begin{figure}
\begin{center}
\includegraphics [width = 100mm]{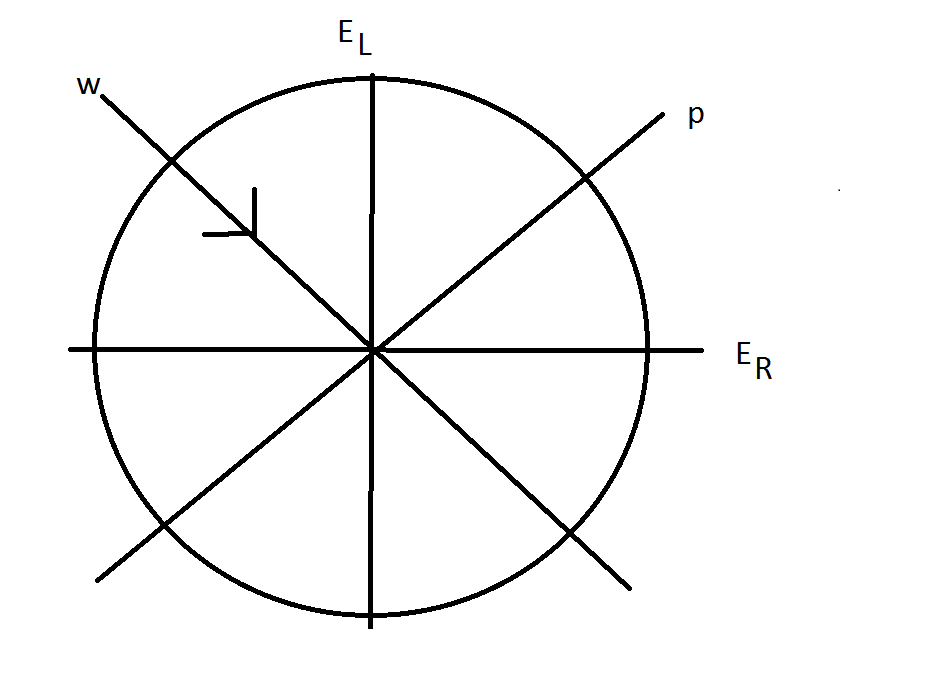}
\caption{Non-heterotic tachyon trajectory}
\end{center}
\end{figure}
 
In the heterotic case, the trajectory is a hyperbola, and there is a second critical temperature (the upper Hagedorn point)  with $\beta _{+} /\pi = \sqrt{2} - 1$  where the trajectory leaves this circle (see Figure 2).\\
\begin{figure}
\begin{center}
\includegraphics[width = 100mm]{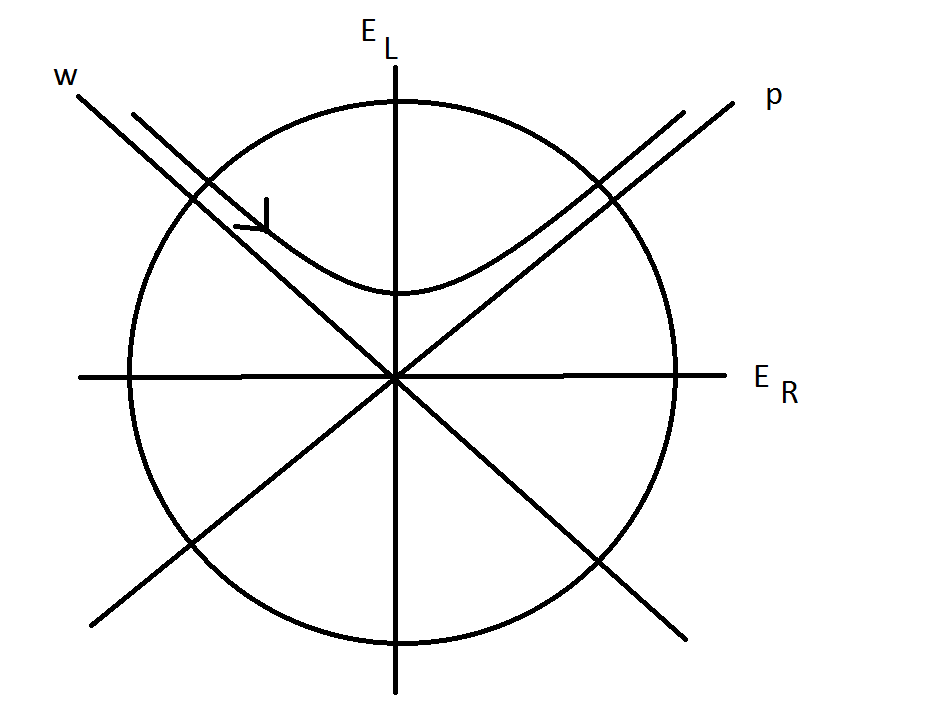}
\caption{Heterotic tachyon trajectory}
\end{center}
\end{figure} 

Similarly to the Higgs mechanism where the tachyon is modified by quartic or higher interactions, thermalon interactions may allow a condensate phase to form at high temperatures. The thermalon would effectively have a positive mass in the condensate phase, so that there would be a finite density of string, string degrees of freedom and entropy in this phase.\\

For an Abelian gauge field $A^{ i} _{ \mu}$ in a heterotic string theory, the right-moving component is represented by the spacetime index $\mu$, while the left component is represented by $i$, a left-heterotic winding direction.\\

A frame field description of gravity is natural  in heterotic string theory, with the  graviton $e^{ i}  _{\mu}$ and gravitino $\psi ^{ \alpha} _ {\mu}$ described  using left-moving world and right-moving frame indices. Consider the timelike component of the metric frame field $e^ 0_ \mu$.
 On Wick rotation in the thermal geometry, the left index $0$ is a winding direction, and the thermalon carries momentum in this direction. At the self-dual point ($T=1/\pi$) the combination of wrapping and momentum contributes only to the left-moving conformal weight. Here the relationship between $e^0_\mu$ and the thermalon field $\phi$ is the same as that between a $U(1)$ gauge field and a charged scalar.\\

Continuation to the imaginary time regime makes like charges for this $U(1)$  repel. This will allow us to make a comparison between the frame-thermalon system and a Ginzburg-Landau superconductor \cite{GinzburgLandau} in the Euclidean regime, which will be relevant in the case where $F$ is near a global minimum.\\

Suppose that we introduce a gravitational gradient, producing a  deviation from Minkowski geometry but with a timelike Killing vector that may be used to define energy and temperature. The effect of the gravitational gradient is to make length of this  Killing vector vary, so that $ \beta$, the local size of compact dimension will vary accordingly. Note that although $\beta$ varies locally, its relationship with the gravitational gradient and Killing vector length means that the temperature viewed by an external observer is constant.\\

We will explore the possibility of both strong and weak thermalon excitations in equilibrium with the vacuum. The Unruh effect \cite{Unruh}, whereby an accelerating Rindler observer measures a nonzero temperature proportional to the acceleration
\begin{equation}
T = \frac{\alpha}{2\pi} 
\end{equation}
indicates that this is possible, as in principle the high temperatures needed for the Hagedorn transition may be obtained from high accelerations. We will show that there are solutions for the heterotic thermalon corresponding to accelerating excitations, which have zero overlap with the horizon. These excitations would represent alternative thermal string ensembles with a nonzero thermalon condensate, in equilibrium with the normal vacuum. These solutions are incompatible with a Minkowski background, but would provide a way to interpolate between concentric exterior Schwarzschild metrics.The possible relevance of these ensembles to gravitational collapse will be explored below. Two (extreme) cases will be examined below in detail, for $\phi$  small compared to the string tension scale, and for $\phi$ in the nonlinear region.

\section{Weak $\phi$ deformations, in Minkowski backgrounds}

In this section, we will investigate the possibility of weak thermalon deformations of Minkowski space. Essentially, we will show that there is an obstruction to the existence of such deformations, due to a warp in the transverse metric across a solution with nonzero $\phi$.  

On gauging away the phase of $\phi$ in the Euclidean regime and continuing to Minkowski space, the Lagrangian for the thermalon takes the form  \cite{AtickWitten} for small $\phi$ 

\begin{equation}
\mathcal{L} = \frac{1}{2}\sqrt{-g}\partial_{\mu} \phi \partial^{\mu} \phi + \frac{1}{2}m^2(\beta)\phi^2 + \frac{1}{4!}\lambda (\beta) \phi^4 + O(\phi ^6)
\end{equation}

where the squared mass $m^2$ is given by 

\begin{equation}
m^2(\beta) = -6 + \frac{\pi^2}{\beta^2} +\frac{\beta^2}{\pi^2}
\end{equation}

The higher order terms in $\phi$ describe the effects of string interactions. In general, the higher order terms cannot be calculated in the absence of a string field theory formalism. However, at the critical points where $m^2(\beta) = 0$ a constant $\phi$ corresponds to on-shell thermalons. Now on-shell, the 4 thermalon scattering amplitude is accessible to perturbation theory, and takes a form similar to the Virasoro-Shapiro amplitude, giving a nonzero value for $\lambda$ which is the same at the two critical points by $T$ duality. The effective $\lambda(\beta)$ between the critical points will be relevant to the issue of thermalon condensation, and as long as this is non-zero a finite condensate will form.\\

In this section we will consider the case of of excitations with small values of $\vert \phi \vert$ in a flat Minkowski background, for which the non-linear terms (which are of higher order in $\vert \phi \vert$) may be neglected (and so the uncertainty regarding their values is not relevant). This we will call the regime of weak excitations. The case for larger values of $\vert \phi \vert$ will be considered below, to build up a picture of how thermalon condensation may progress during gravitational collapse. \\

In this regime we will seek solutions which appear static to a Rindler observer, which can be in thermal equilibrium with the surrounding vacuum at a position dependent Unruh temperature. It should be emphasised that we do not take such accelerated observers to have a privileged status beyond the accelerated excitation solutions taking a simple form in such frames. From the standpoint of general covariance, in such a frame we can take $\phi$ to be a scalar, although it should be remembered that this is essentially a shorthand for the entity $(\phi,e^0,\beta)$ or equivalently $(\phi, \beta e^0)$, as $e^0$ is needed to define the local temperature and Wick rotation. If $E_0$ is the vector field dual to $e^0$, then the local energy and temperature are defined relative to the Killing vector $\beta E_0$.\\

Note that in Minkowski space the choice of $E_0$ is arbitrary due to Lorentz invariance.  However, the conjecture that we will develop below is that thermalon deformations of the ordinary vacuum are only relevant where $E_0$ is parallel to a timelike Killing vector, and in the case of interest here, of near spherically symmetric collapsed objects, this will be essentially unique.\\

Using Rindler coordinates $(\tau, r, x, y)$ \cite{BirrellDavies} for such an accelerated observer, the inverse temperature $\beta$ depends on the distance $r$ from the horizon as $\beta = \frac{2\pi}{\alpha} = 2\pi r$, and the effective squared mass of the thermalon $m^2$ varies accordingly, as

\begin{equation}
m^2(r) = (\frac{1}{4r^2} +4r^2 -6)
\end{equation}

For thermal equilibrium $\partial_{\tau} \phi$ = 0, and for a plane wave solution $\partial_x \phi =\partial_y \phi = 0$, and the differential equation for $\phi$ becomes
\begin{equation}
\frac{\partial^2 \phi}{\partial r^2} + \frac{1}{r}\frac{\partial \phi}{\partial r} = m^2(\beta (r)) \phi = (\frac{1}{4r^2}+ 4r^2 -6)\phi \label{DE}
\end{equation}

Near the horizon, $r \rightarrow 0$, we have $m^2 \sim (1/4r^2 - 6)$ and (\ref{DE}) simplifies to a Bessel form and the solution which is finite at $r=0$ has

\begin{equation}
\phi \sim J_{1/2}(r) \sim \sqrt{r}
\end{equation}

Likewise it is easy to verify that $\phi$ will have a Gaussian cut-off at large $r$. In fact, the exact solution of (\ref{DE}) which is finite at $r=0$ is

\begin{equation}
\phi =\epsilon \sqrt{r} \exp (-r^2) \label{sol}
\end{equation}
where $\epsilon$ is an arbitrary small parameter. Here $\epsilon$ has the same dimensions as $\phi$ (energy), and the solution is shown in Figure 3. The simple form of $\phi$ will enable us to evaluate various integrals of interest below.\\
\begin{figure}
\begin{center}
\includegraphics [height = 50mm, width = 100mm]{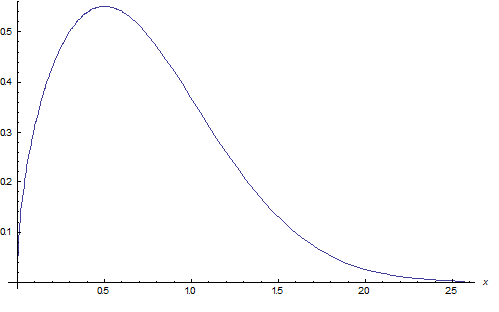}
\end{center}
\caption {$\phi / \epsilon$ for the weak heterotic solution}
\end{figure}

Note that for this solution, $\phi$ takes its maximum value at $r=1/2$ where $\beta = \pi$, the self-dual value. The solution for $\phi$ vanishes at the horizon, and the solution is essentially confined to $r<2.5$, as $\phi$ diminishes rapidly beyond this point.\\

For comparison, let us compare this solution to the situation for a non-heterotic thermalon, for example that for the bosonic string. Here,  \cite{AtickWitten}

\begin{equation}
M^2 = -8+\frac{\beta^2}{\pi^2}=-8+4r^2
\end{equation}
and the solution which is finite for $r \rightarrow 0$ has

\begin{equation}
\phi = cJ_{0} ~2\sqrt(2)r)
\end{equation}
which tends to a non-zero value as $r \rightarrow 0$ and the exact solution, with $\phi(0) = \epsilon$, may be given in terms of a Laguerre polynomial of fractional order as

\begin{equation}
\phi = \epsilon \exp(-x^2)L_{1/2}
\end{equation}

Whereas this solution has the same Gaussian cut-off at large $r$ as for the heterotic case, the thermalon excitation is not now excluded from the horizon. Thus thermal duality can be seen to be responsible for confining the weak thermalon excitation solution away from the horizon (the other independent solution diverges at horizon and is unphysical) which would be relevant in the environment of a gravitationally collapsed object. The heterotic thermalon excitation is thus kept from falling into the horizon. An `excitation trap' effect is thus present in this case.

\section{Back reaction on the metric}

Although the solution presented above was calculated using a special reference frame, the excitation would have properties such as the distortion of the metric produced by back reaction on which all observers would agree. This back reaction will be seen to produce a transverse warp, presenting an obstruction to embedding in a MInkowski background.\\

The thermalon excitation solution in the previous section apparently accelerates through Minkowski spacetime. This would be mechanically possible if the overall inertia of the solution vanished, with positive mass on the forward side, negative mass on the following side and positive pressure in between acting against the inertia (positive and negative) on either side. We will now look in more detail at the back reaction of the thermalon excitation `wall' solution on the metric. It will turn out that the excitation has a positive overall gravitational mass. Acceleration of a configuration with positive mass through Minkowski space is clearly impossible, as the conservation of momentum and energy would be violated. Rather than inhabiting a conventional Minkowski background, it will be shown that the thermalon excitation wall interpolates between two Minkowski regions with a transverse warp factor between them. \\

The Einstein field equations are given by

\begin{equation}
R_{\mu \nu}-\frac{1}{2}g_{\mu \nu}=8\pi GT_{\mu \nu}
\end{equation}
where we have retained Newton's constant $G$ as we are working in string tension units. We will work to first order in $\epsilon$, so that the solution above remains valid. Introducing notation for the $\phi$ kinetic term

\begin{equation}
K_{\mu \nu} = \partial_{\mu} \phi \partial_{\nu} \phi
\end{equation}
the Einstein equations become, on implementing the thermal equilibrium constraints

\begin{equation}
\delta(\beta^2 g^{00}) = 0, K_{00} = 0
\end{equation}
and re-arranging, as follows (where V is the poential):

\begin{equation}
\frac{R_{00}}{4\pi G} = g_{00}( V-\frac{1}{2}\frac{\partial V}{\partial \ln \beta}) \label{RZZ}
\end{equation}

\begin{equation}
R_{11} = R_{00} +8\pi G( K_{11} - V) \label{R11}
\end{equation}
and similarly for $R_{22}$ and $R_{33}$.\\

Now from (\ref{RZZ}) we have a local gravitational source $\mu = \rho +\ Tr( p)$ given by, with $\theta = \beta^{-2} = T^2$

\begin{equation}
\mu = \frac{1}{2}\frac{\partial ( \theta V)}{\partial \theta}
\end{equation}
and introducing $y = r^{-2}$ gives

\begin{equation}
V =\frac{1}{2} (\frac{1}{4}y +4 y^{-1}- 6)\phi^2
\end{equation}
and now

\begin{equation}
\mu = \frac{1}{2}\frac{\partial (\theta V)}{\partial \theta} = \frac{1}{2}\frac {\partial(yV)}{\partial y} = \frac{1}{2}(\frac{1}{4 r^2} - 3)\phi^2
\end{equation}
as

\begin{equation}
y = 4 \pi^2 \theta
\end{equation}
Introducing a factor of $\beta / \beta_0$ to compensate for the redshift seen by an observer at the arbitrary location $r_0 = \beta_0/2\pi$ gives

\begin{equation}
\frac{\beta \mu}{\beta_0} = \pi e^{-2r^2}(\frac{1}{4}-3r^2)\epsilon^2/\beta_0
\end{equation}
and 

\begin{equation}
\frac{1}{\beta_0}\int_{0}^{\infty}\beta \mu dr = \frac{1}{2\beta_0}(\frac{\pi}{2})^{3/2}\epsilon^2
\end{equation}
which is finite and positive. Note that $\mu$ is zero at

\begin{equation}
r = \frac{1}{2\sqrt{3}}
\end{equation}
and is negative to the high temperature side of this point and positive to the low temperature side.\\

Now the warp factor across the wall can be calculated by an application of the equivalence principle. If an area $A$ behind the wall matches an area $A' = A + \delta A$ ahead of the wall, and we define the warp factor $w$ by $w-1 = \delta A/A$ then with $g$ the acceleration due to gravity in the Rindler frame,
\begin{equation}
\beta_1 g_1 A_1- \beta_0 g_0 A_0 = 4\pi G \int_V \beta \mu dv \sim  4\pi GA\int \beta \mu dr
\end{equation}
where the volume $V$ extends across the comparison surface and between $r_0$ and $r_1$. Now letting $r_0 \rightarrow 0$ and $r_1 \rightarrow \infty$ gives

\begin{equation}
2\pi \delta A \sim 4\pi GA \int_{0}^{\infty}\mu \beta dr = 2\pi GA(\frac{\pi}{2})^{3/2}\epsilon^2
\end{equation}
and finally

\begin{equation}
w-1 = G (\frac{\pi}{2})^{3/2}\epsilon^2
\end{equation}
This warp factor is realised by a hyperbolic spatial geometry within the thermalon `wall', which is another aspect of the back reaction, described by (\ref{R11}). This hyperbolic geometry is a feature that the weak thermalon wall solution shares with the strong $\phi$ solutions discussed below.\\

Although the plane wave solution is not viable in an ordinary Minkowski space background, where conservation of energy requires a counter-factual warp factor,  this analysis does suggest the possibility of a standing wave solution in a strong gravitational field, which we will examine next.

\section{Weak $\phi$ deformations, in a Schwarzschild background}

The exterior Schwarzschild metric is given by

\begin{equation}
ds^2 = (1-\frac{R_s}{R})dt^2 - (1-\frac{R_s}{R})^{-1}dR^2 - R^2(d\theta^2 + \sin^2\theta d\phi^2) 
\end{equation}
where $R_s$ is the Schwarzschild radius, $R_s = 2GM$ at which there is an event horizon. For large $M$, we will assume that the background is sufficiently flat that the Hagedorn transition points are not significantly affected; the results below will be qulitatively similar even if these points are perturbed a little by the background. We will use this exterior Schwarzschild metric, but will work in a region excluding the horizon. We will argue that in fact a closed horizon will never form due to thermalon effects. \\

For a thermalon deformation to be a viable solution, it must interpolate between concentric (approximately) Schwarzschild solutions, with the thermalon wrapping aligned with (the Wick rotation of) a timelike Killing vector to give a standing thermalon excitation just outside the virtual horizon. For a large black hole, the surface is almost flat and the plane wave thermalon solution given above will locally be a good approximation to the true solution. To leading order in $1/M$, the area of a static surface at  the self-dual temperature $T = 1/\pi$ will differ from the area of the horizon by a fixed amount (independent of $M$) of $O(1)$, due to the approximately hypercylindrical form of the spatial metric. The area difference  $\delta A$ that comes from the warp factor $w$ of the static thermalon excitation will therefore provide an interpolation to another Schwarzschild metric to the exterior. The  horizon area that would be predicted from the exterior solution is increased by $\delta A$ compared to the actual horizon area. This shows the possibility of thermalons shrinking the actual horizon area for a given mass. We will argue below that in fact a closed horizon will never form as the energy present will be converted to thermalon form during collapse. Whereas this altered black hole solution can be seen as a deformed version of the conventional black hole, in the scenario below the black hole is replaced completely by a region of thermalon condensate. \\ 

The warp factor of the thermalon solution means that thermalon deformations of the Minkowski vacuum are not possible. Only in the near horizon geometry will coherent and persistent thermalon excitations be possible.  The persistence of strongly accelerated thermalon traps in these solutions is the key to the possibility of persistent deformations, rather than simply the presence of curvature (which is small for high mass backgrounds).\\

Our basic conjecture is that a particle of energy $m$ (measured from infinity) falling through a Schwarzschild geometry will begin to nucleate thermalons when the geometry to its outside locally acquires a timelike Killing vector with critical acceleration. Thereafter, this condition will be maintained by kinetic energy being fed from the particle to increase the thermalon density in the static trap.  This process is thermodynamically favoured because of the large amount of entropy released, which we will show in the next section will be esentially the Hawking entropy for an equivalent black hole.

 \section{Hawking thermodynamics and the thermalon}

The basis of Hawking's analysis of the entropy of black holes is the relationship
\begin{equation}
dS = \beta dQ = \beta dM \label{S}
\end{equation}
where $\beta$ is the vacuum temperature, which expresses the principle that energy emitted by a black hole is completely thermalised. Since the thermalon excitation is in thermal equilibrium with the vacuum, this relationship should also apply to our solution

\begin{equation}
\phi = \epsilon \sqrt r \exp (-r^2) \label{sol2}
\end{equation}
interpreted either as a static or a dynamic excitation. 
Now (\ref{S}) may be expressed in terms of the free energy as
\begin{equation}
d F = 0
\end{equation}
or equivalently

\begin{equation}
d \ln (Z) = 0
\end{equation}
Now $\ln Z$ consists of a kinetic part and a potential part, which can be evaluated for the solution (\ref{sol}) by direct integration. First, the kinetic part per unit area is given by

\begin{equation}
\int_{0}^{\infty}\beta \frac{1}{2}(\frac{\partial \phi}{\partial r})^2 dr
\end{equation}
and inserting the solution (\ref{sol2}) we have
\begin{equation}
\int_{0}^{\infty}2\pi r \frac{1}{2}(\frac{\partial}{\partial r}(r\exp(-r^2))^2 dr\epsilon^2 = \frac{\pi}{4}\sqrt{\frac{\pi}{2}}\mathrm{erf}(\infty)\epsilon^2=\frac{1}{2}(\frac{\pi}{2})^{3/2}\epsilon^2
\end{equation}
The potential part of $ \ln Z$ per unit area  is

\begin{equation}
\int_{0}^{\infty}\beta V dr
\end{equation}
and using
\begin{equation}
V = \frac{1}{2}(\frac{1}{4r^2}+4r^2 -6)\phi^2
\end{equation}
we have

\begin{equation}
V =  \frac{1}{2}(\frac{1}{4r^2}+4r^2 -6)r\exp(-2r^2)\epsilon^2
\end{equation}
and
\begin{equation}
\beta V = \pi (\frac{1}{4} + 4r^4 -6r^2)\exp(-2r^2) \epsilon^2
\end{equation}
giving

\begin{equation}
\int_{0}^{\infty}\beta V dr = -\frac{1}{2}(\frac{\pi}{2})^{3/2}\epsilon^2
\end{equation}
so that 

\begin{equation}
\delta \ln Z = 0
\end{equation}
confirming the thermal nature of the excitation. For a static excitation bounded by the horizon we will have a differential form of the Hawking area law
\begin{equation}
\delta S=\frac{\delta A}{4G}
\end{equation}
\\

\section{Conversion mechanism}

The process of conversion of kinetic to thermalon energy will proceed from the outside inwards. Due to the large release of entropy in thermalon condensation outlined above, the motion of an infalling particle is essentially determined by the condition that the static surface gravity is regulated by thermalon condensation. The kinetic energy of the particle reduces with the enclosed area to mainatin a constant static surface gravity, and the thermalon warp increases to compensate, so that the thermalon region continues to interpolate between the Schwarzschild solutions to the inside and outside of the particle. Note that the term involving the time derivative of $\phi$  in the thermalon action will be suppressed by a factor $O(\ln{Mm})^{-2}$ compared to the other terms. \\

 The distance to fall for a particle of mass $m$ between the start of thermalon nucleation and merger with an existing collapsed object of mass $M$ is $O(\sqrt {Mm})$, and during this the kinetic energy is reduced from $Mm$ to $O(1)$  (note the squared reduction factor, so that the particle undergoes a double deceleration compared to the local surface gravity).   \\

Although the particle spectrum is Lorentz invariant, the thermalon modes that they couple to during the conversion process are not, as the possibility of thermalon deformations is dependent on the formation of static traps, which are not Lorentz invariant.\\ 

Assume that the wavefunction for a particle of mass $m$ is spread out transverely over an area $A$. The change of the horizon area on adding the mass $m$ is $2Mm$. and we assume  that $A \gg Mm$ (so that there is a small relative area change). The area change for thermalon parameter $\phi$ over the area $A$ scales like $A \phi ^2$, so that $A \phi^2 \sim  Mm$ and $\phi^2 \ll 1$, and the weak thermalon regime applies, so that a consideration of higher order terms in $\phi$ is not necessary in dealing with the particle conversion process. \\

For the plane wave thermalon solution to be relevant, it is only necessary that the infalling particle wavefunction be spread out on a scale large compared to $\sqrt{M}$, which is roughly of nuclear dimensions for a stellar mass collapsing object, and of atomic scale for a galactic core object. If the infalling particle cannot be assumed to be delocalised, then the following observation is relevant. During its fall from the initiation of thermalon condensation to an existing thermalon region, through a distance of  $O(\sqrt{mM})$, the thermalon Green's function can spread over an area $O(mM)$, comparable to the increase in surface area of the condensate region, so that the weak $\phi$ limit is valid for infalling particles through most of their trajectory.\\

The production of thermalons due to the deceleration of the nucleating particle may arise through a Larmor like radiation process. The radiated power would in that case scale as $\sim (ma)^2$ which will equal $ma$ for relativistic motion, so that $ma \sim 1$. Interactions between thermalons and light particles that could generate thermalon bremsstrahlung are accessible on shell through conformal field theory, and a self-consistent interaction berween light particles and thermalons  which would support the conversion process may be possible.\\ 

It may be noted that the interaction between matter falling through the excitation region is similar to that of particles moving through a detector e.g. a cloud chamber. In both cases a dissipative process converts part of the kinetic energy of the particle into a trace carrying information about the particle's trajectory.\\

A freely falling observer, on passing into the conversion region, would also be subject to these processes, and would effectively be converted into a Rindler observer at a string tension scale acceleration and temperature and thus would experience an encounter with a kind of `firewall' in the conversion region. \\

\section{High temperature string phase}
Once a closed static trap has formed within the collapsing body, the kinetic energy of infalling particles will undergo thermalon conversion and increase the value of $\phi$ there.  As $\phi$ strengthens, it is driven towards the non-linear region, where the small $\phi$ approximation used above no longer applies. Eventually a stationary point in the $\phi$ potential is approached and  a phase transition should occur. The region where $\phi$ approximates the stationary value will thereafter expand, rather than the value of $\phi$ increasing further.  We will first examine the bulk properties of this phase by looking for a space filling solution with constant $\phi$. For this space-filling solution, a low energy approximation is valid for the bulk region because there the space and time derivatives of $\phi$ are small. \\ 

If we assume the  presence of higher order terms in the thermalon potential, there will be a minimum for a given $\beta$. Further, for the heterotic models there will be a space filling solution with constant $\beta$ and $\phi$. Let $(\beta_s , \phi_s)$
be the point in $(\beta, \phi)$ space where $T^2 V$ has its global minimum. This will exist between the upper and lower Hagedorn temperatures, by the mean value theorem. Then from (\ref{RZZ}) we have

\begin{equation}
R_{00} = \frac{1}{2}\frac{\partial (\tau V)}{\partial \tau} = 0
\end{equation}
and also

\begin{equation}
\frac{\partial V}{\partial \phi} = \frac{1}{\tau}\frac{\partial (\tau V)}{\partial \phi} = 0
\end{equation}
allowing a solution with $(\beta, \phi) = (\beta_s, \phi_s)$. 
For $\phi$ constant in space, and with $R_{00} = 0$ we have from (\ref{R11})

\begin{equation}
R_{11} = - 8\pi G V_s g_{11}
\end{equation}
where $V_s = V(\beta_s, \phi_s)$ and similarly for $R_{22}$ and $R_{33}$, giving a hyperbolic spatial geometry.
 \begin{equation}
du^{2} =  \frac{dr^{2}}{1+a^{-2}r^{2}} + r^{2}(d \theta^{2} +
\sin^{2} \theta d \phi^{2}) \label{BL}
\end{equation}
and a space-time interval
\begin{equation}
ds^{2} =  dt^{2} - du^{2}
\end{equation}
with
\begin{equation}
a^{-2} = -8\pi G V > 0
\end{equation}\\
It can be seen that the spatial geometry must undergo a radical radial deflation during the conversion to the high temperature string phase.\\

In the effective theory with real $\phi$, if a gravitational gradient is introduced, it will produce a gradient in $\beta$. Any gradient of $\beta$ around $\beta = \beta_{s}$ will produce a polarization of $\mu$ with positive source on the high $\beta$ side and negative source on the low $\beta$ side since $z$ has a maximum at the stable point $\beta = k\pi$. This provides the gravitational London current required to expel gravitational gradients (up to some breakdown point). The London response is given by 
\begin{equation}
\frac{\partial \mu}{\partial \beta} 
\end{equation} 
which changes sign at $\beta_s$.\\

We conjecture that finite regions of this gravitationally superconducting phase can exist, with an outer boundary layer with a positive gravitational source, interpolating between the high temperature string phase and the surrounding vacuum, with which it will be in thermal equilibrium.  Any gravitational source in the exterior is effectively screened so that it appears only in this boundary layer. \\

As well as the thermalon condensate, the high temperature phase will be expected to contain a finite density of string excitations, due to the positive mass of the thermalon mode around the ground state. Because of the short range of gravity, which is effectively restricted to the penetration depth, the gravitational field of these strings is screened, allowing them to exist without generating event horizons. This indicates a basic role for the thermalon mechanism in maintaining dynamical consistency for the heterotic string models, as otherwise most string states would be hidden by event horizons \cite{Hewitt93}. The string spectrum is thereby restored, with long strings existing in an alternative phase in a way reminiscent of colour confinement in QCD. As explained below, the emergence of these long string excitations in the condensate phase opens the possibility of an entanglement between long and short string sectors being set up during collapse and conversion, which may allow a reconciliation between the generation of the Hawking entropy and the principles of quantum mechanics.\\

Due to the hyperbolic geometry of the high temperature string phase interior, the volume of the bulk is proportional  to the surface area for high mass. This makes the bulk thermodynamic properties of the high temperature string phase region, and particular the entropy, consistent with the area scaling of the Hawking entropy. The Hilbert space for such a region might correspond to a field theory on a `fuzzy sphere' \cite{Madore}, \cite{t'Hooft}. A stringy collapsed state without a horizon or singularity was proposed in \cite{LuninMathur} as a resolution of the black hole information paradox.\\ 

During a collapse, a shell region of the high temperature string phase once formed  will expand both outwards, as infalling  matter reaches the outer surface of the shell drawing an extended region of thermalon condensation in behind it, and inwards as the contracting inner surface of the shell engulfs matter to the inside.\\ 

In a realistic collapse, concentric shells may form, but these will merge as the collapse progresses. These processes will continue until a stable, near spherical configuration is attained, with the interior filled with the high temperature string phase.\\ 

The boundary region of a shell is almost completely transparent as the thermalon condensate does not carry electric, colour or other charges. We conjecture that the effective opacity builds with shell thickness, as the excitation strings hosted by the shell will interact with normal matter. The combination of exponential extinction with the logarithmic dependence of the shell  thickness with area ratio indicates that the resulting transmission coefficient should be a power of $A_{inner}/A_{outer}$, the ratio of its outer to inner surface areas. As the inner surface of the shell shrinks, the limit of a black body is approached.\\

Overall, the conversion process of kinetic energy to thermalon excitations to high temperature string phase  provides a dissipative mechanism by which the energy released by gravitational collapse produces a hot final state without horizons or singularities. The entropy of this final state, relative to an observer to whom only short string states are available for measurement, matches the Hawking entropy, which is here generated during the braking process.. The large Hawking entropy in this context means that the relatively small number of ordered collapsing states will end up in final states whose bulk properties are the same as those of a very much larger number of other possible states in the high temperature string phase.\\ 

Note that the  thermalon has no supersymmetry partner, so the screening mechanism gives mass to the $e^0$ component of gravity, but not to the gravitino,  illustrating that thermal effects do not respect supersymmetry.\\

\section{Microscopic  interpretation of the dissipative entropy}
Reconciling the process of gravitational collapse with quantum mechanics and quantum information theory presents two problems: what happens to the information stored in the collapsing body, and where does the very large amount of apparently random information produced during the Hawking evaporation process come from? In our scenario, the first problem is easily resolved, as there is no singularity, and the infalling matter is simply transformed into another form. However, our analysis of the thermalon indicates that the high temperature string phase will have esentially the same large area dependent entropy as in the Hawking case. What is the microscopic description of this entropy, and how can it arise in a way consistent with quantum mechanics?\\

Entropy is related to incomplete information, or access to only part of a complete system. In quantum mechanics, this can be expressed through the concept of entanglement entropy. If the state of the system under consideration is ideally isolated from its envionment, we may write

\begin{equation}
|T> = |S>|E>
\end{equation}
where $|T>$ represents the total state in Hilbert space, $|S>$ the  state of the system and $|E>$ the state of the environment. In general, however there will be entanglement between the system and environment so that

\begin{equation}
|T> = M_{ij}|S^i>|E^j>
\end{equation}
and if only the system $S$ is accessible to measurement, it has an effective density matrix

\begin{equation}
\rho_i^k = M_{ij}<E^j|E_l>M^{*lk}
\end{equation}
and our incomplete knowlwdge of the pure state $|T>$ is quantified by the entropy $S = \mathrm{Tr} \rho \log \rho$.\\

The thermalon entropy calculated in thermodynamic terms above should have a microscopic quantum description, and in particular the large thermalon entropy generated during the conversion process should be consistent with the preservation of a pure overall state $|T>$. Before the conversion process begins, the short and long string sectors are essentially decoupled, and the long string sector is in the ground state so that

\begin{equation} \label{F}
|T> = |S>|0_L>
\end{equation}
Now if we divide space into concentric interior, shell and exterior regions we can write

\begin{equation}
|S> = |S^I_i>|S^S_j>|S^E_{kl}>M^{ik}M^{jl}
\end{equation}
where $i,j,k,l$ represent multiple tensor indices related to a fuzzy sphere related decomposition \cite{Madore}, \cite{t'Hooft} into angular momentum modes, with the order of $i$ proportional to the inner surface area of the shell and $j$ to the difference between outer and inner surface areas - the orders of $k$ and $l$ together are proportional to the outer surface area. If a similar decomposition is made for $|0_L>$, the conversion process will be represented by a unitary transfromation $U$ on the states $|S^S_j>$ and $|0_{LJ}^S>$.  The factorisation $(\ref{F})$ no longer holds, and the short string sector now has an entanglement entropy proportional to the difference in surface area between the inside and outside of the shell. This will increase as the conversion process proceeds, eventually reaching a value proportional to the outer surface area. This provides a mechanism by which the short string (field) sector can acquire an area related entanglement entropy which, for consistency with the thermodynamic analysis above, should coincide with the Hawking entropy.   \\

The long strings would provide the heat bath in the interior, which would interact with photons etc. (via biquadratic and higher order interactions) to maintain a temperature consistent with continuity with the external vacuum.  \\

As the entropy apparently generated by the conversion process is mainly due to a conversion of entanglement, the form of the entropy is approximately universal and its thermal nature, as shown in the production of (the equivalent of) Hawking radiation, has its origin in the entanglement entropy of the near (virtual) horizon region.  In this way, the evolution of the system does not require the generation of a large amount of new random information, and can be consistent with the general principles of quantum mechanics.\\

 The thermalon scenario provides an information buffer for information to escape from the interior which gives a kind of ‘firewall’ that will not fall into the hole, as there is a physical mechanism to support it against collapse. In the final state, the interior space has been crushed into a hyperbolic geometry, effectively a hologram only a few hundred Planck lengths thick which holds all the information from the collapsed object and the entanglement partners of Hawking photons.\\

\section{Timescales}

In this section we will identify various timescales associated with the conversion process outlined above, and make estimates for the width of the dynamic trap region for typical astrophysical scenarios of terminal gravitational collapse.\\

The timescale for external accumulation, in which the remaining outer part of the collapsing body undergoes conversion after the initiation of the conversion processl is $O(M \ln M)$, and the time taken (as measured by an external observer) for the conversion process to form a complete hyperbolic internal region is also $O(M \ln M)$. This is much shorter than the time taken for the collapsed object to decay by Hawking like radiation, which will be $O(M^3)$. The average rate at which string would be produced within a collapsing object against the clock of an external observer would have an average value of $O(1)$ independent of the mass $M$ of the object. \\

The time taken for the distribution of string around a shell to relax to a uniform configuration will vary from $O(M)$ to $O(M^2)$ depending on the degree of anisotropy of the infalling matter.\\

For a ten solar mass object, we have $M \sim  2$x$10^{31} $kg, and the $\sqrt{M}$ scale of the dynamic trap region is equivalent to $\sim 5$x$10^{-16}$m, while a ten million solar mass object has  $\sqrt{M} \sim 5$x$10^{-13}$m.

\section{Conclusions}

In this paper we have identified an accelerating excitation (the`thermalon excitation') of the heterotic string vacuum, which could propagate consistently in a spacetime background of the kind that would be found in gravitational collapse, allowing the formation of a  configuration without a closed horizon. \\

There appears to be a viable mechanism by which the energy released in extreme gravitational collapse may be converted to thermalons, which would accumulate to produce a region of a high temperature string phase in the $E(8)$ x $E(8)$ and $SO(32)$ string models. This would provide away of braking the motion of infalling matter and leading to an endpoint free of event horizons or singularities.\\

A possible resoultion of the apparent conflict between quantum mechanics and the generation of the Hawking entropy through an entanglement between the short and long string sectors has been proposed. \\

The mechanism proposed may also give other benefits in confirming the dynamical consistency of string theory by making the string spectrum stable against horizon formation. Bearing in mind the fictitious nature of the thermalon, and the dependence of thermal effects upon the observer, this may be summarised as proposing that the resolution of the horizon problem raised in \cite{Hewitt93} is that the consistency of string theory is related to the emergence of coherent string tower effects in strong gravitational fields. \\

As no event horizon is formed, the problems of cross horizon entanglement arising from conventional gravitational collapses are avoided \cite{Almheiri}.\\ 

The scenario presented here would also avoid the radical violation of time reversal symmetry that is involved in admitting conventional black holes but not their time reversed counterparts.\\

Although we have concentrated on the Schwarzschild case in 4 dimensions, similar results should apply without spherical symmetry and in different numbers of uncompactified dimensions.  In particular, the shell formation mechanism does not rely on strict spherical symmetry and should be a robust and generic feature of extreme gravitational collapse in heterotic string models.\\

\section{Acknowledgements}
The author would like to thank Dan Waldram for useful discussions, and Nick Mavromatos and Steven Duplij for helpful comments and suggestions.\\

\end{document}